\documentclass[a4paper,12pt]{article}
\usepackage{epsfig}
\usepackage{times}
\usepackage{graphicx}
\usepackage{citesupernumber}
\usepackage{naturefem}
\usepackage{nature}
\usepackage{amsmath}
\usepackage{color}
\ifx\pdfoutput\undefined
 \DeclareGraphicsExtensions{.epsi,.eps}
\else
 \DeclareGraphicsExtensions{.pdf}
\fi

\newcommand{\mycite}[1]{\cite{#1}}

\newcommand{\beq}{\begin{equation}}
\newcommand{\eeq}{\end{equation}}
\newcommand{\beqa}{\begin{eqnarray}}
\newcommand{\eeqa}{\end{eqnarray}}
\newcommand{\ket} [1] {\vert #1 \rangle}
\newcommand{\bra} [1] {\langle #1 \vert}

\def\ket#1{|\,#1\,\rangle}
\def\bra#1{\langle\, #1\,|}

\def\opone{\leavevmode\hbox{\small1\kern-3.8pt\normalsize1}}

\newcommand{\eins}{\mbox{$1 \hspace{-1.0mm} {\bf l}$}}
\begin{document}


\baselineskip24pt \noindent{\Large\bf Experimental Device-independent Tests of Classical and Quantum Dimensions}\\[3mm]
\noindent {\bf Johan Ahrens$^1$, Piotr Badzi{\c a}g$^1$, Ad\'an Cabello$^{2,1}$, \& Mohamed Bourennane$^1$}\\
\noindent {\it $^{1}$Physics Department, Stockholm University, S-10691 Stockholm, Sweden.}\\[2mm]
\noindent {\it $^{2}$Departamento de F\'{\i}sica Aplicada II, Universidad de Sevilla, E-41012 Sevilla, Spain.}\\[2mm]




\noindent \textbf{A fundamental resource in any communication and computation task is the amount of information that can be transmitted and processed. Information encoded in a classical system is limited by the dimension $d_c$ of the system, i.e., the number of distinguishable states. A system with $d_c=2^n$ classical states can carry $n$ bits of classical information. Information encoded in a quantum system is limited by the dimension $d_q$ of the Hilbert space of the system, i.e., the number of perfectly distinguishable quantum states. A system with $d_q=2^n$ perfectly distinguishable quantum states can carry $n$ qubits of quantum information. Physical systems of higher dimensions may enable more efficient and powerful information processing protocols. The dimension is fundamental in quantum cryptography and random number generation, where the security of many schemes\mycite{ACM06,PB11,LYWZWCGH11} crucially relies on the system's dimension. From a fundamental perspective, the dimension can be used to quantify the non-classicality of correlations, since classical simulation of correlations produced by a quantum system of dimension $d_q$ may require a classical system of dimension $d_c \gg d_q$\mycite{KGPLC10,CJ11,Cabello12}. For all these reasons, a fundamental problem in information theory is to assess the (classical or quantum) dimension of a physical system in a ``device-independent'' scenario, i.e., without referring to the system's specifications, which may be under control of a dishonest supplier, eavesdropper or saboteur. In this contribution we report experiments realizing this goal for systems emitted by a black box. Our results indicate that dimension witnesses utilized in the experiments may become a powerful tool for testing systems provided by unreliable sources.}


\section*{Introduction}

\begin{figure}[h]
\centerline{\includegraphics[width=14cm]{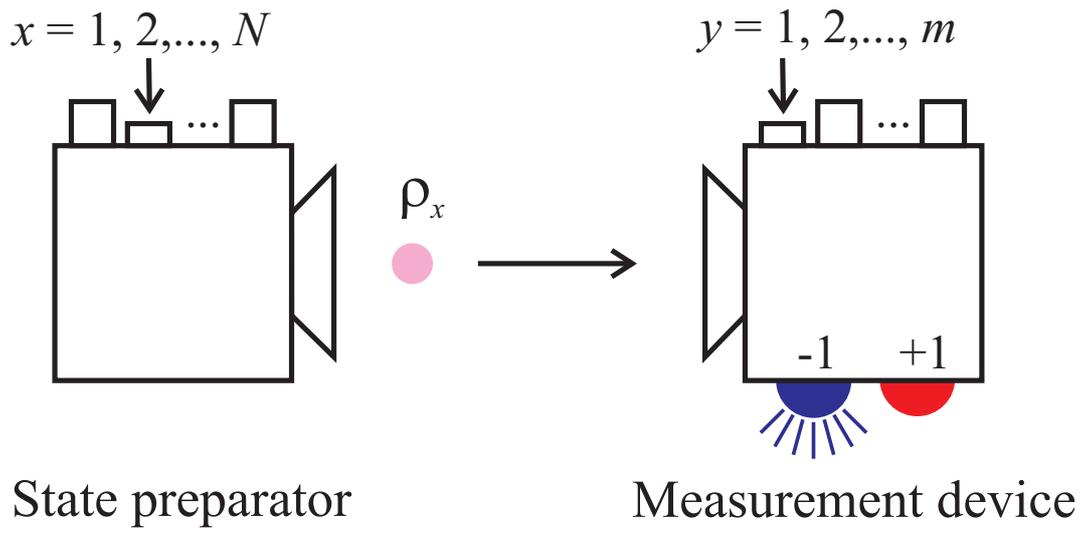}}
\caption{\label{Fig1}Device-independent scenario for testing the minimum classical or quantum dimension.}
\end{figure}

The problem of testing the minimum dimension of a system has been considered from different theoretical perspectives. Brunner \emph{et al.}\mycite{BPAGMS08} introduced the concept of a quantum dimension witness for the dimension of the Hilbert space of composite systems tested locally. Wehner, Christandl and Doherty\mycite{WCD08} related this problem to the construction of quantum random access codes. Wolf and P\'erez-Garc\'{\i}a\mycite{WP09} approached the question from a dynamical viewpoint. Finally, Gallego \emph{et al.}\mycite{GBHA10} addressed the problem from a ``device-independent'' viewpoint, without any reference to the type of system (e.g., quantum or composite), its evolution, preparation or the measurement devices. The device-independent scenario is illustrated in Fig. \ref{Fig1}. There is a state preparator, a black box, which has $N$ buttons and emits a system in a (classical or quantum) state $\rho_x$ when button $x\in\{1,\ldots,N\}$ is pressed. The prepared state is then sent to the measurement device, another black box, which has $m$ buttons and produces an outcome $b \in \{-1,+1\}$ when button $y\in\{1,\ldots,m\}$ is pressed. The results of the test are characterized by the probability distribution $P(b|x,y)$ for obtaining result $b$ in measurement $y$ on state $\rho_x$. Within this scenario, tight classical dimension witnesses and quantum dimension witnesses can be experimentally used to determine the minimum classical and quantum dimension of the system produced by the state preparator and measured by the measurement device. These dimension witnesses are combinations of probabilities $P(b|x,y)$ and the upper bounds for their maximum values depend on the system's dimension.

Here we report several experimental tests of sets of states supplied by a potentially contaminated source (a black box from the tester's point of view). Each time the goal is to identify lower bounds for $d_c$ and $d_q$ of systems supplied by the black box. In each experiment the black box emits physical systems prepared in a different way. The experimental results are characterized by the expectation values
\begin{equation}
 \label{mean}
 E_{xy}= P(+1|x,y)-P(-1|x,y),
\end{equation}
which are the quantities needed for the dimension witnesses. Specifically, we considered two combinations of expectation values introduced by Gallego \emph{et al.}\mycite{GBHA10}, called $I_3$ and $I_4$, to test the dimension of the systems.

The first combination, $I_3$, works both as a tight two-dimensional classical witness and a two-dimensional quantum witness. It uses three preparations ($N=3$) and two dichotomic measurements ($m=2$). The corresponding inequalities are:
\begin{equation}
 \label{I3}
 I_3 \equiv | E_{11}+E_{12}+E_{21}-E_{22}-E_{31} | \stackrel{\text{bit}}\leq 3 \stackrel{\text{qubit}}\leq 1 + 2 \sqrt{2} \stackrel{\text{trit,qutrit}}\leq 5,
\end{equation}
where $\stackrel{\text{bit}}\leq 3$ means that no classical system of dimension $d_c=2$ can give a value larger than 3, and $\stackrel{\text{qubit}}\leq 1 + 2 \sqrt{2}$ means that no quantum system of dimension $d_q=2$ can give a value larger than $1 + 2 \sqrt{2} \approx 3.8284$. Finally, $\stackrel{\text{trit, qutrit}}\leq 5$ means that no classical system of dimension $d_c=3$ or quantum system of dimension $d_q=3$ can give a value larger than 5, which is the algebraic maximum of $I_3$.

The second case, $I_4$, requires four preparations ($N=4$) and three dichotomic measurements ($m=3$). It represent several witnesses. Their underlying inequalities are
\begin{equation}
 \label{I4}
 I_4 \equiv E_{11} + E_{12}+ E_{13} + E_{21} + E_{22}- E_{23} + E_{31} - E_{32} - E_{41} \stackrel{\text{bit}}\leq 5 \stackrel{\text{qubit}}\leq 6 \stackrel{\text{trit}}\leq 7 \stackrel{\text{qutrit}}\leq 2 + \sqrt{13 + 16\sqrt{2}} \stackrel{\text{quart,ququart}}\leq 9.
\end{equation}
Thus $I_4$ is a dimension witness for $d_c=2,3$ and $d_q=2,3$.


\section*{Experimental setup}


The state preparator in Fig.~\ref{Fig2} emits single photons in which information is encoded in horizontal ($H$) and vertical ($V$) polarizations, and in two spatial modes ($a$ and $b$). We define four basis states: $|0\rangle \equiv |H,a\rangle$, $|1\rangle \equiv |V,a\rangle$, $|2\rangle \equiv |H,b\rangle$ and $|3\rangle \equiv |V,b\rangle$. With these encodings, any qubit state can be represented as $\alpha |H,a\rangle + \beta |V,a\rangle$, any qutrit state as $\alpha |H,a\rangle + \beta |V,a\rangle + \gamma |H,b\rangle$, and any ququart state as $\alpha |H,a\rangle + \beta |V,a\rangle + \gamma |H,b\rangle + \delta|V,b\rangle$. The single photon source in Fig.~\ref{Fig2} emits horizontally polarized photons. The states needed for our experiments are produced by suitably adjusting the three half wave plates $\theta_i$, HWP($\theta_1$), HWP($\theta_2$) and HWP($\theta_3$), where $\theta_i$ is the rotation angle of the corresponding plate, in the setup of the state preparator in Fig.~\ref{Fig2}. The general form of the prepared state is
\begin{equation}
 \begin{split}
 \ket{\psi} = &
 \sin{(2\theta_1)}\cos({2\theta_2)}|H,a\rangle +
 \sin{(2\theta_1)}\sin{(2\theta_2)}|V,a\rangle\\
 &+\cos{(2\theta_1)}\cos{(2\theta_3)}|H,b\rangle +
 \cos{(2\theta_1)}\sin{(2\theta_3)}|V,b\rangle.
 \end{split}
\end{equation}
For those experiments in which the state preparator emits qubit states, $P(+1|x,y)$ is obtained from the number of detections in $D_1$, and $P(-1|x,y)$ is obtained from the number of detections in $D_3$.
For those experiments in which it emits qutrit states, $P(+1|x,y)$ is obtained from the number of detections in $D_1$ and $D_2$, and $P(-1|x,y)$ is obtained from the number of detections in $D_3$.
For those experiments in which the prepared states are trit or ququart states, the base states are mapped directly to one detector each, as follows: $|0\rangle \rightarrow D_1$, $|1\rangle \rightarrow D_3$, $|2\rangle \rightarrow D_2$ and $|3\rangle \rightarrow D_4$.

For the experiments where $I_3$ and $I_4$ reach the algebraic limit, the prepared states $\ket{\psi_i}$ are the eigenstates of $I_3$ and $I_4$, respectively, and they are mapped to detectors $D_i$, where $i=1,\ldots,3$ for $I_3$, and $i=1,\ldots,4$ for $I_4$. All these mappings are done by adjusting half wave plates HWP($\varphi_1$) and HWP($\varphi_2$) in the setup of the measurement device in Fig.~\ref{Fig2}. The single photons were emitted from a $780$ nm diode laser. The laser was attenuated so that the two-photon coincidences were negligible. Our four single-photon detectors were Silicon avalanche photodiodes calibrated to have the same detection efficiency. All single counts were registered using a four-channel coincidence logic unit with a time window of $1.7$ ns. This logic unit associates each count of the detectors to an outcome $-1$ or $+1$. The number of detected photons was about $2.10^4$ per second. The measurement time for each experiment was $30$ s.


\section*{Experimental results}


We performed experiments to test the minimum dimension of systems emitted by a state preparator.

The first experiment is an $I_3$ test of a system of qubits. The goal is to obtain the maximum qubit violation of the bit bound $I_3(d_c=2)=3$. For this purpose, we prepared $N=3$ qubit states and performed $m=2$ dichotomic measurements which maximize the value of $I_3$. The optimal states and measurements for all experiments are described in Methods. The goal of the second experiment is to obtain the maximum qutrit violation of the qubit bound of $I_3(d_q=2)\approx 3.8284$. For this, we prepared $N=3$ qutrit states and performed $m=2$ dichotomic measurements which reach the algebraic bound $I_3=5$.

All other experiments are $I_4$ tests. The goal of the third experiment is to obtain the maximum qubit violation of the bit bound $I_4(d_c=2)=5$. For this purpose, we prepared $N=4$ qubit states and performed $m=3$ dichotomic measurements which maximize $I_4$. The goal of the fourth experiment is to obtain the maximum trit violation of the qubit bound $I_4(d_q=2)=6$. For this, we prepared $N=4$ trit states and performed $m=3$ dichotomic measurements which maximize $I_4$. The goal of fifth experiment is to obtain the maximum qutrit violation of the trit bound $I_4 (d_q=3) =7$. For this, we prepared $N=4$ qutrit states and performed $m=3$ dichotomic measurements which maximize $I_4$. The sixth experiment is an $I_4$ test on ququarts. The goal is to obtain the maximum ququart violation of the qutrit bound $I_4 (d_q=4) =7.96887$. For this, we prepared $N=4$ ququart states and performed $m=3$ dichotomic measurements which reach the algebraic bound $I_4=9$.

The states which saturate the witness' boundaries may not be valuable for information processing. It is therefore interesting to test the dimension for states which are useful for information processing purposes. For quantum cryptography, a valuable set of states consists of four pairwise orthogonal and pairwise unbiased qubit states, such as $\ket{\psi_{1}} = \ket{0}$, $\ket{\psi_{2}}=\frac{1}{\sqrt{2}}(\ket{0} + \ket{1})$, $\ket{\psi_{3}}=\ket{1}$ and $\ket{\psi_{4}}=\frac{1}{\sqrt{2}}(\ket{0}-\ket{1})$\mycite{BB84}. The seventh experiment consists of violating the bit bound of $I_4 (d_c=2)=5$ using these four cryptographic states. We performed the measurements maximizing the value of $I_4$ for these states (see Methods). These measurement settings give $I_4 (BB84)=\sqrt{2} + 2 + \sqrt{5} \approx 5.6506$, which clearly exceeds the bit bound $I_4 (d_c=2)=5$.

All experimental results are shown in Fig. \ref{Fig3}. These results are in very good agreement with the theoretical values and demonstrate that we are able to determine the minimum dimension of the emitted states. The small errors were due to imperfections in the optical interferometers, the non perfect overlapping and coupling of the light modes, and the polarization components. The error bars were deduced from the propagated Poissonian counting statistics of the raw detection events.


\section*{Conclusions}


We have experimentally determined lower bounds for the dimension of several ensembles of physical systems in a device-independent way. We tested classical and quantum dimension witnesses derived by Gallego \emph{et al.}\mycite{GBHA10} For this purpose, we prepared photonic qubits, trits, qutrits and ququarts in optimal states and performed optimal measurements to maximally violate the corresponding dimension witnesses. In addition, we measured a dimension witness on the four qubit states used in standard quantum cryptography. Our results demonstrate how dimension witnesses can be utilized to test classical and quantum dimensions of physical systems supplied by unreliable sources and distinguish between classical and quantum states of a given dimension. A very good agreement between the experimental results and the theoretical predictions makes us believe that the method can be extended to more complex witnesses and to tests of systems claiming to span higher dimensions.


{\bf Acknowledgements} The authors thank A.\ Ac\'{\i}n, E.\ Amselem and R.\ Gallego for stimulating discussions. This work was supported by the Swedish Research Council (VR), the Linnaeus Center of Excellence ADOPT, the MICINN Project No.\ FIS2008-05596 and the Wenner-Gren Foundation.\\


\section*{Methods}


\noindent{\bf Maximum qubit violation of the bit bound of $I_3$ }To design $N=3$ qubit states and $m=2$ dichotomic measurements which maximize the value of $I_3$, we consider the two dichotomic measurements
\begin{equation}
 \label{measurements}
 M_k = \eins - 2\ket{m_k}\bra{m_k},
\end{equation}
where $\eins$ denotes the identity matrix and
\begin{equation}
 \label{vectors}
 \ket{m_{1,2}} = \cos\left(\frac{x}{2}\right)\ket{0} \mp \sin\left(\frac{x}{2}\right)\ket{1}.
\end{equation}
The three prepared states can be chosen as pure states $\rho_k = \ket{\psi_k}\bra{\psi_k}$, where
\begin{equation}
 \ket{\psi_3} = \cos\left(\frac{x}{2}\right)\ket{0} - \sin\left(\frac{x}{2}\right)\ket{1} = \ket{m_1}.
\end{equation}
The optimization of the setup can thus be reduced to finding the maximum of the sum of the larger eigenvalues of $M_1 + M_2$ and $M_1 - M_2$. This fixes parameter $x$ to $x_{\rm opt} = \frac{\pi}{4}$ with the following result:
\begin{equation}
 \ket{m_{1,2}} = \frac{\sqrt{2 + \sqrt{2}}}{2}\ket{0} \mp \frac{\sqrt{2 - \sqrt{2}}}{2}\ket{1}.
\end{equation}
States $\ket{\psi_1}$ and $\ket{\psi_2}$ are then the corresponding eigenvectors, $\ket{\psi_1} = \ket{1}$ and $\ket{\psi_2} =\frac{1}{\sqrt{2}}(\ket{0} + \ket{1})$.


\noindent{\bf Maximum qutrit violation of the qubit bound of $I_3$ }To reach the algebraic bound $I_3=5$ with qutrits, we used
\begin{subequations}
\begin{align}
 M_1 &= \ket{0}\bra{0} - \ket{1}\bra{1} + \ket{2}\bra{2},\\
 M_2 &= \ket{0}\bra{0} + \ket{1}\bra{1} - \ket{2}\bra{2},
\end{align}
\end{subequations}
and the states $\ket{\psi_1} = \ket{0}$, $\ket{\psi_2} = \ket{2}$ and $\ket{\psi_3} = \ket{1}$.


\noindent{\bf Maximum qubit violation of the bit bound of $I_4$ }To determine the maximum qubit violation of $I_4$, we generalize the procedure used for $I_3$. We consider three measurements $M_k$ ($k=1,2,3$), with $\ket{m_1}$ and $\ket{m_2}$ defined in \eqref{measurements} and \eqref{vectors}. State $\ket{m_3}$ is arbitrary. The optimization of the setup is now reduced to maximizing the sum of the largest eigenvalues of $M_1-M_2$, $M_1+M_2+M_3$ and $M_1+M_2-M_3$. It brings the optimal value of $x$ to $x_{\rm opt} = \frac{\pi}{6}$ and $\ket{m_3} = \frac{1}{\sqrt{2}}(\ket{0} - \ket{1})$. The corresponding states are then the eigenvectors belonging to the maximal eigenvalues of $M_1+M_2+M_3$, $M_1+M_2-M_3$, $M_1-M_2$ and $-M_1$, i.e.,
\begin{subequations}
\begin{align}
 \ket{\psi_1} &= (2 + \sqrt{3})\ket{0} + \ket{1},\\
 \ket{\psi_2} &= (2 + \sqrt{3})\ket{0} - \ket{1}, \\
 \ket{\psi_3} &= \ket{0} + \ket{1}, \\
 \ket{\psi_4} &= \ket{m_1}.
\end{align}
\end{subequations}


\noindent{\bf Maximum trit violation of the qubit bound of $I_4$ }The optimal preparations are
$|\psi_1\rangle = |\psi_2\rangle=|0\rangle$, $|\psi_3\rangle = |2\rangle$ and $|\psi_4\rangle = |1\rangle$, and
the optimal measurements are
\begin{subequations}
\begin{align}
 M_1 &= |0\rangle \langle 0|-|1\rangle \langle 1|+|2\rangle \langle 2|,\\
 M_2 &= |0\rangle \langle 0|+|1\rangle \langle 1|-|2\rangle \langle 2|,\\
 M_3 &= |0\rangle \langle 0|-|1\rangle \langle 1|-|2\rangle \langle 2|.
\end{align}
\end{subequations}


\noindent{\bf Maximum qutrit violation of the trit bound of $I_4$ }The optimal measurements correspond to the observables of the form \eqref{measurements} with
\begin{equation}
\ket{m_{1,2}} = \cos\left(\frac{x}{2}\right)\ket{1} \pm \sin\left(\frac{x}{2}\right)\ket{2}.
 \end{equation}
The optimization proceeds as for qubits, but the algebra is more involved. We obtain
\begin{subequations}
\begin{align}
 \ket{m_3} &= \frac{1}{\sqrt{2}}(\ket{0} + \ket{2}), \\
 \ket{\psi_4} &= \ket{m_1}, \\
 \ket{\psi_3} &=\frac{1}{\sqrt{2}}(\ket{1} - \ket{2}),
\end{align}
\end{subequations}
and the (unnormalized) $\ket{\psi_2}$ and $\ket{\psi_1}$
\begin{equation}
 \ket{\psi_{2,1}} =\ket{0} \pm \left[1 - \cos{x} - \sqrt{1 +
 (1-\cos{x})^2}\right]\ket{2}.
\end{equation}
The optimal value of $\cos{x}$ is now $\cos{x_0} = \frac{1}{2}(1- \sqrt{2} + \sqrt{2\sqrt{2} -1}) = 0.4689$. It gives $I_4 (d_q=3) = 2 + \sqrt{13 + 16\sqrt{2}} = 7.9688$.


\noindent{\bf Maximum ququart violation of the qutrit bound of $I_4$ }The optimal preparations are
$|\psi_1\rangle = |0\rangle$, $|\psi_2\rangle = |2\rangle$,
$|\psi_3\rangle = |1\rangle$ and $|\psi_4\rangle = |3\rangle$, and
the optimal measurements are
\begin{subequations}
\begin{align}
 M_1 &= |0\rangle \langle 0|+|1\rangle \langle 1|+|2\rangle \langle 2|- |3\rangle \langle 3|,\\
 M_2 &= |0\rangle \langle 0|-|1\rangle \langle 1|+|2\rangle \langle 2|-|3\rangle \langle 3|,\\
 M_3 &= |0\rangle \langle 0|+|1\rangle \langle 1|-|2\rangle \langle 2|-|3\rangle \langle 3|.
\end{align}
\end{subequations}


\noindent{\bf Violation of the bit bound of $I_4$ with cryptographic states }The measurement settings maximizing the value of $I_4$ for the standard cryptographic states are specified by the vectors
\begin{subequations}
\begin{align}
 \ket{m_1} &= \ket{\psi_4}, \\
 \ket{m_2} &= (c \sqrt{1-p} - s \sqrt{p})\ket{0} + (c \sqrt{p} + s \sqrt{1-p})\ket{1}, \\
 \ket{m_3} &= \frac{c-s}{\sqrt{2}}\ket{0} + \frac{c+s}{\sqrt{2}}\ket{1},
\end{align}
\end{subequations}
where $c=\cos\left(\frac{\pi}{8}\right)$, $s=\sin\left(\frac{\pi}{8}\right)$ and $p=\frac{1}{2} \left(1+\frac{3}{\sqrt{10}}\right)$. These measurement settings give $I_4 (BB84)=\sqrt{2} + 2 + \sqrt{5} \approx 5.6506$.



\newpage


\newpage


\begin{small}
\begin{figure}[h]
\centerline{\includegraphics[width=15.5cm]{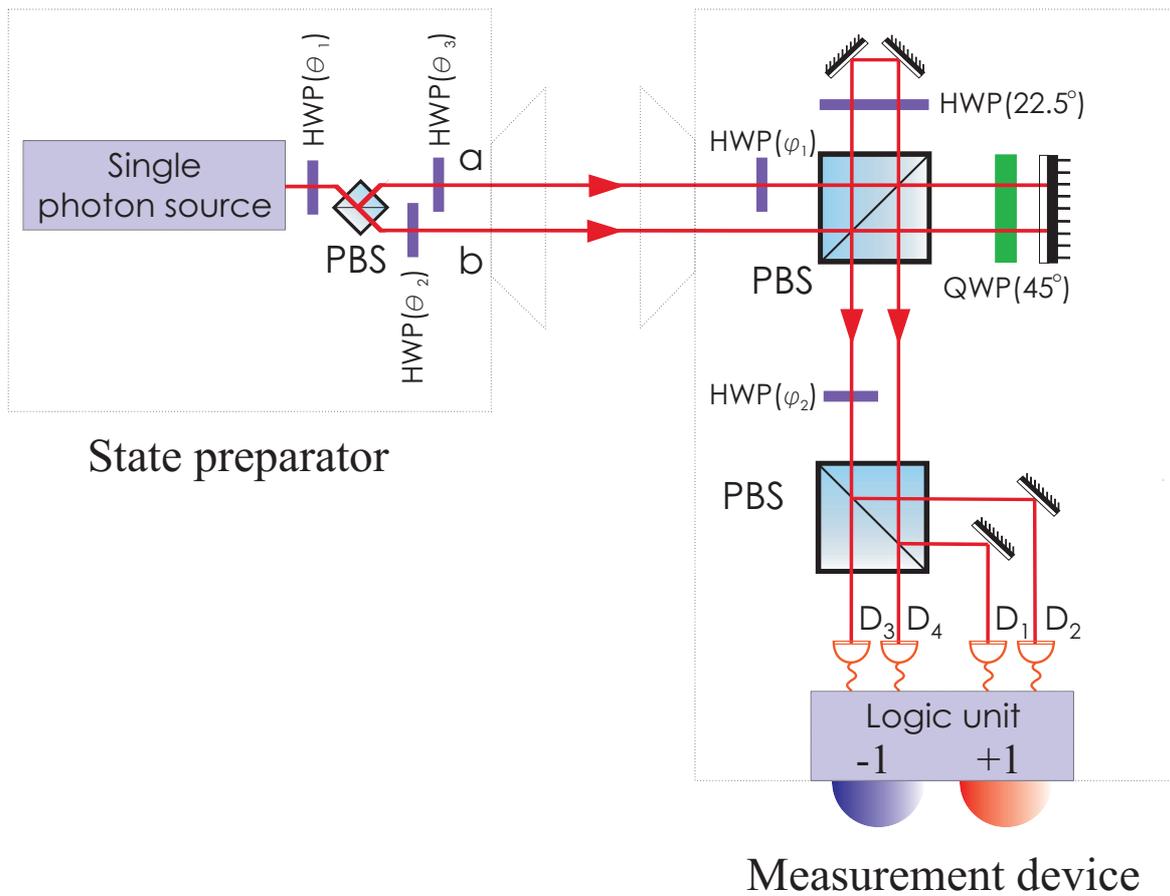}}
\caption{\label{Fig2}Experimental setup for testing classical and quantum dimension witnesses.}
\end{figure}
\end{small}


\newpage


\begin{small}
\begin{figure}[h]
\centerline{\includegraphics[width=13cm]{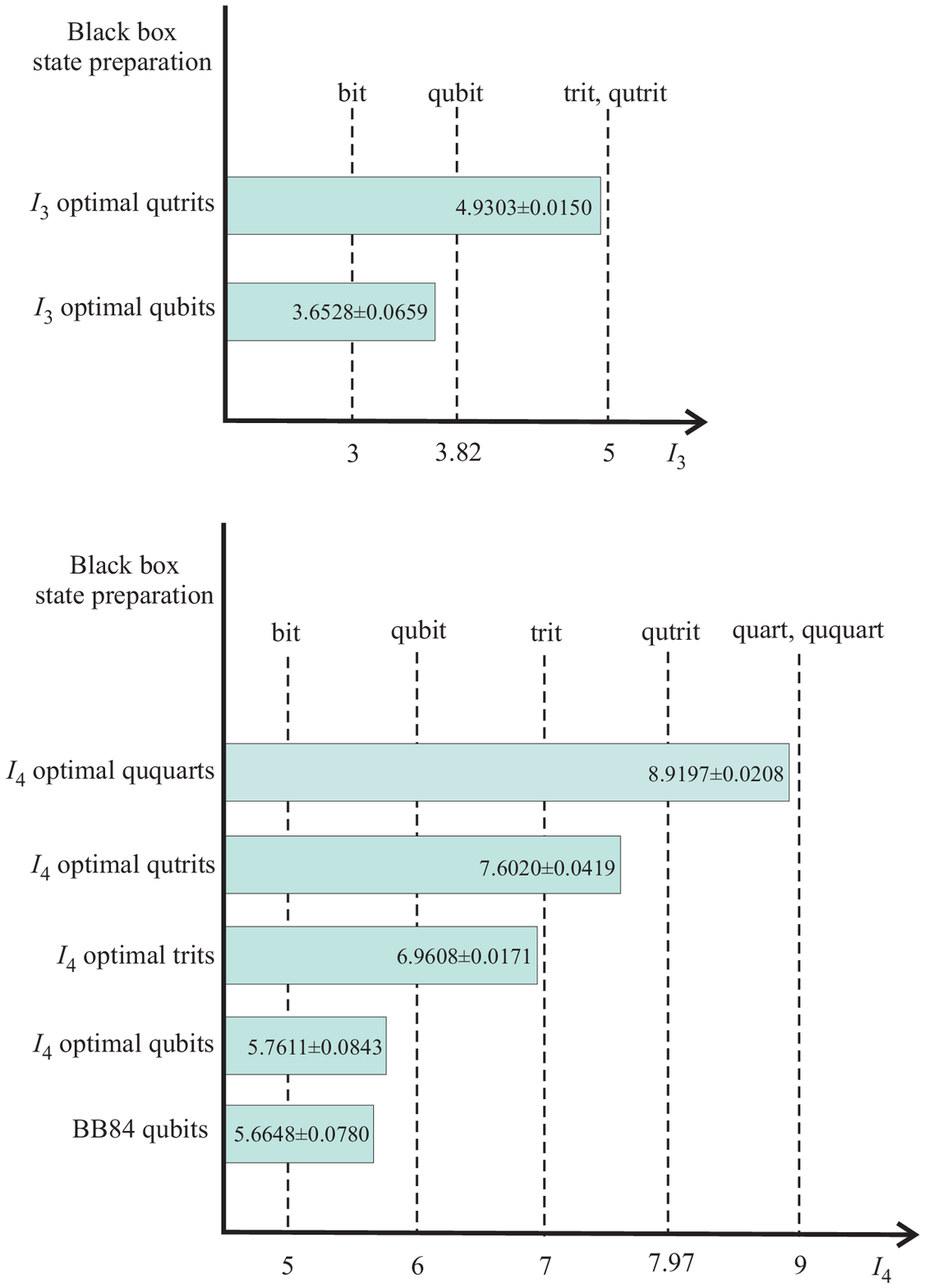}}
\caption{\label{Fig3}Experimental results of the dimension witness tests. The vertical dashed line labeled ``bit'' represents the maximum value achievable with bits, and similarly for the other vertical dashed lines. ``$I_3$ optimal qutrits'' means that the black box actually emits qutrit states which give the maximum value for $I_3$ using qutrits, and similarly for the other preparations. ``BB84 qubits'' denotes the states used in standard quantum cryptography.}
\end{figure}
\end{small}



\begin{thebibliography}{19}


\bibitem{ACM06}
 Ac\'{\i}n, A., Gisin, N. \& Masanes, L.
 From Bell's theorem to secure quantum key distribution.
 \emph{Phys. Rev. Lett.} \textbf{97} 120405 (2006).

\bibitem{PB11}
 Paw{\l}owski, M. \& Brunner, N.
 Semi-device-independent security of one-way quantum key distribution.
 \emph{Phys. Rev. A} \textbf{84}, 010302(R) (2011).


\bibitem{LYWZWCGH11}
 Li, H.-W. \emph{et al.}
 Semi-device-independent random-number expansion without entanglement.
 \emph{Phys. Rev. A} \textbf{84}, 034301(R) (2011).


\bibitem{KGPLC10}
 Kleinmann, M., G{\"u}hne, O. Portillo, J. R., Larsson, J.-{\AA} \& Cabello, A.
 Memory cost of quantum contextuality.
 \emph{New. J. Phys.} (2011).

\bibitem{CJ11}
 Cabello, A. \& Joosten, J. J.
 Hidden variables simulating quantum contextuality increasingly violate the Holevo bound.
 In \emph{Unconventional Computation},
 edited by Calude, C. S., Kari, J., Petre, I. \& Rozenberg, G.
 \emph{Lecture Notes in Computer Science} \textbf{6714} (Springer, Berlin, 2011), 64--76.

\bibitem{Cabello12}
 Cabello, A.
 The contextual computer.
 In \emph{A Computable Universe},
 edited by Zenil, H.
 (World Scientific, Singapore, 2012).


\bibitem{BPAGMS08}
 Brunner, N., Pironio, S., Ac\'{\i}n, A., Gisin, N., M\'ethot, A. \& Scarani, V.
 Testing the dimension of Hilbert spaces.
 \emph{Phys. Rev. Lett.} \textbf{100}, 210503 (2008).

\bibitem{WCD08}
 Wehner, S., Christandl, M. \& Doherty, A. C.
 Lower bound on the dimension of a quantum system given measured data.
 \emph{Phys. Rev. A} \textbf{78}, 062112 (2008).

\bibitem{WP09}
 Wolf, M. M. \& P\'erez-Garc\'{\i}a, D.
 Assessing quantum dimensionality from observable dynamics.
 \emph{Phys. Rev. Lett.} \textbf{102}, 190504 (2009).

\bibitem{GBHA10}
 Gallego, R., Brunner, N., Hadley, C. \& Ac\'{\i}n, A.
 Device-independent tests of classical and quantum dimensions.
 \emph{Phys. Rev. Lett.} \textbf{105}, 230501 (2010).


\bibitem{BB84}
 Bennett, C. H. \& Brassard, G.
 Quantum key distribution and coin tossing.
 In {\em Proceedings of IEEE International Conference on Computers, Systems, and Signal Processing, Bangalore, India} (IEEE, New York, 1984), 175--179.

\end{thebibliography}
\end{document}